\documentclass[aps,pra,superscriptaddress,twocolumn,showpacs]{revtex4}
\usepackage[colorlinks, linkcolor=blue, citecolor=blue, urlcolor=blue, breaklinks=true]{hyperref}
\usepackage{amsmath}
\usepackage{graphicx}
\usepackage{amssymb}
\usepackage{bbm, color}
\usepackage{relsize}

\newcommand*{\tn}[1]{{\textnormal{#1}}}

\begin{document}

\title{Activation of quantum capacity of Gaussian channels }
\author{Youngrong Lim}
\affiliation{Department of Mathematics and Research Institute for Basic Sciences, Kyung Hee University, Seoul 02447, Korea}
\author{Soojoon Lee}
\affiliation{Department of Mathematics and Research Institute for Basic Sciences, Kyung Hee University, Seoul 02447, Korea}
\affiliation{Centre for the Mathematics and Theoretical Physics of Quantum Non-Equilibrium Systems, School of Mathematical Sciences, The
 University of Nottingham, University Park, Nottingham NG7 2RD, United Kingdom}

\date{\today}

\begin{abstract}
  Quantum channels can be activated by a kind of channels whose quantum capacity is zero. This activation effect might be useful to overcome noise of channels by attaching other channels which can enhance the capacity of a given channel. In this work, we show that such an activation is possible by specific positive-partial-transpose channels for Gaussian lossy channels whose quantum capacities are known. We also test more general case involving Gaussian thermal attenuator whose the exact value of quantum capacity has been unknown so far. For a recently suggested narrow upper bound on quantum capacity of the thermal attenuator, we confirm the fact that an activation of quantum capacity occurs as well. This result is applicable for realistic situations in which Gaussian channels describe the noises of communication systems.
\end{abstract}

\maketitle

\section{Introduction} \label{intro}
  Quantum channels describe the situation in which any physical process acts on a quantum system~\cite{Shannon}. Quantum channel capacities quantify the maximal amount of information reliably transmitted via noisy quantum channels after performing suitable error correction schemes~\cite{Lloyd, Devetak}. In general channel capacities are not easily calculable owing to regularization of infinitely many uses, and for the quantum capacity, which is the ability of transmitting quantum information, i.e., qubits, through given channels, its explicit calculation has been even less known.
  
  What quantum channels are crucially different with classical channels is that they are not additive in general. Moreover, the quantum capacity of quantum channels has a more striking property, {\it superactivation}, in which the combined capacity of two channels is positive even though each channel has zero capacity~\cite{Science}. This contradicts the intuition that any combination of useless communication channels always gives only another useless channel. 
  
  In practice, a special kind of quantum channels, called Gaussian channels, are considered for many reasons. They can not only well approximate continuous-variable quantum systems in physical situations, but can be also very useful for implementations in a sense that Gaussian states, operations and measurements can be realized by basic elements in quantum optics, i.e., laser, beam splitter, squeezer, etc.~\cite{RMP}. There are seemingly analogous results in between discrete and Gaussian quantum information theory, but no clearcut correspondence exists yet~\cite{Lim16}.
  
 Therefore, a natural step forward is seeking the exotic properties, i.e., (super)activation, in Gaussian channels likewise. Indeed, a superactivation effect has been found for Gaussian channels by Smith, Smolin, and Yard~\cite{Smith11}. They have used a kind of positive-partial-transpose (PPT) channel, i.e., an entanglement binding channel, and a 50\% attenuation channel whose quantum capacities are both zero, but the combined coherent information can be positive. Since coherent information is a lower bound on the quantum capacity, thus we can confirm the superactivation effect although we don't know the exact value of quantum capacity for the combined channel.
 
 In this work, we present more general cases in which one of the channels can have a positive quantum capacity. In particular, since we know the quantum capacity of that channel, we can check whether there is an activation of quantum capacity between them by computing a specific coherent information of the combined channel. In other words, we here show that the combined channel can have higher quantum capacity than sum of the two. We figure out that this is the case when one is an entanglement binding channel and the other is a Gaussian lossy channel whose transmissivity is slightly above 50\%. 
 
Furthermore, we apply our method to the thermal attenuator, which is more general version of the Gaussian lossy channel~\cite{upper,upper4}. From our result, it turns out that the coherent information of the whole channel also exceeds the upper bound on quantum capacity of  the thermal attenuator for specific range of parameter, which shows the occurrence of activation.
 
 In section~\ref{review}, we briefly review quantum capacity and Gaussian channels. We present our main results in section~\ref{main}, and conclude our work with further remark in section ~\ref{conclusion}.
 
 \section{Background} \label{review}
 
 A quantum channel $\mathcal{N}_{A\rightarrow B}:\rho_A \rightarrow \rho_B$ is defined by a completely positive trace preserving (CPTP) map from a quantum state $\rho_A$ on $\mathcal{H}_A$ to another quantum state $\rho_B$ on $\mathcal{H}_B$. The quantum capacity of the given quantum channel is the capability of reliably transmitting quantum information through the channel. Then we have a useful lower bound on quantum capacity of a channel, called the coherent information,
 \begin{equation} \label{coh}
 I_c=H(B)-H(E)
 \end{equation}
where $H$ is the von Neumann entropy defined as $H(\rho)=-\tn{Tr}~\rho ~\tn{log}_2 \rho$. $H(B)$ means the von Neumann entropy of output state, i.e., $H(\mathcal{N}_{A\rightarrow B}(\rho_A))$, and $H(E)$ is the von Neumann entropy of environment, which can be calculated from the entropy of the complementary channel, $H(\mathcal{N}^c_{A\rightarrow B}(\rho_A))$. Then the one-shot quantum capacity is the maximum value of the coherent information over all input state $\rho_A$ such 
as $\mathcal{Q}^{(1)}=\underset{\rho_A}{\max}~I_c$, and the quantum capacity of a given channel $\mathcal{N}$ is written as
\begin{equation} \label{qcapacity}
\mathcal{Q}(\mathcal{N})=\lim_{n\rightarrow \infty} {\mathcal{Q}^{(1)}(\mathcal{N}^{\otimes n}) \over n },
\end{equation}
where $n$ is the number of parallel channel uses. Owing to the fact that these maximization and infinite regularization are hard to calculate in general, the exact values of quantum capacities have been known only for a few special cases. However, it is enough to consider the coherent information Eq.~(\ref{coh}) only for some inputs which can show such a (super)activation occurs since $I_c \leq \mathcal{Q}^{(1)} \leq \mathcal{Q}$.

\begin{figure}[!t]
\includegraphics[width=9.3cm]{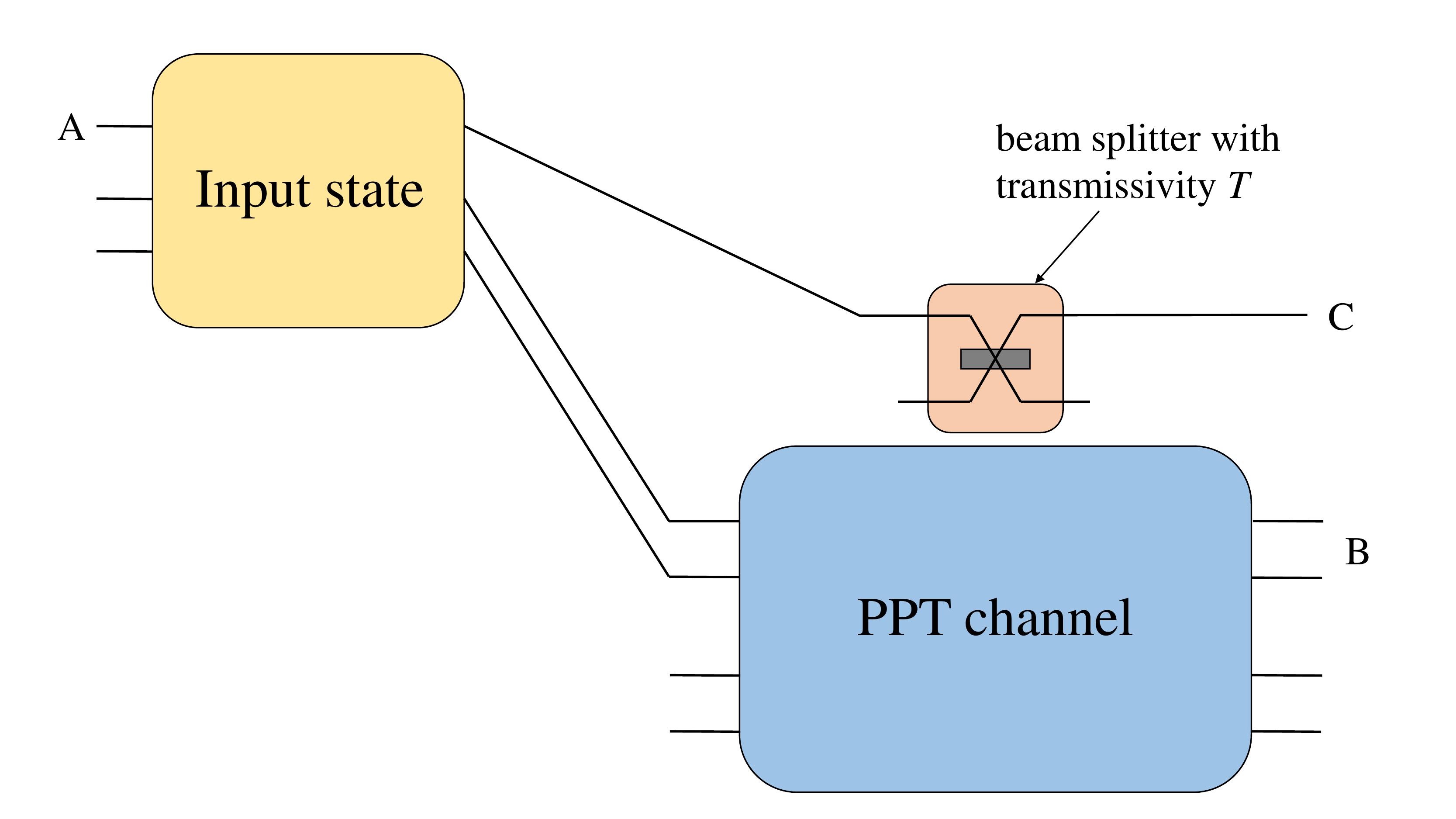}
\caption{A schematic diagram of the overall channel process. The mode A, thermal input, goes through the combined channel after appropriate purification. The output of channel is three modes, B and C. Remaining modes are the environment assumed to be vacuum.}
\label{fig1}
\end{figure}

In case of Gaussian channels, however, we should consider Gaussian states in an infinite dimensional Hilbert space. Instead directly applying a channel operation on a Gaussian state itself, we can rather consider a covariance matrix $\gamma$, a $2n \times 2n$ real symmetric matrix for an $n$-mode Gaussian state, which carries all the crucial information of the Gaussian state~\cite{Holevo}. Then the action of the Gaussian channel on this covariance matrix can be expressed as
\begin{equation}\label{Gch}
\gamma \rightarrow X\gamma X^t + Y,
\end{equation}
where $X$ and $Y$ are matrices satisfying the condition $Y+i(J-XJX^t) \geq 0$ with $J=\begin{pmatrix} 0 & 1 \\ -1 & 0 \end{pmatrix}^{\oplus n}$, and $X^t$ is the transpose of $X$. Indeed, these matrices are related with the symplectic operation $S$ of the Gaussian channel. When a Gaussian channel sends input $A'$ to output $B$, there is its symplectic dilation with ancillary input mode $E'$ and output environment $E$, similarly to the unitary dilation for the discrete-variable case~\cite{Wilde}. Then the channel operation is described by the symplectic matrix $S$ such as
\begin{equation} \label{Symplectic}
S=\begin{pmatrix} X & Z \\ X_c & Z_c \end{pmatrix},
\end{equation}
where $X$ is the same matrix as in Eq.~(\ref{Gch}) and $Y=ZZ^t$. $S$ is symplectic in a sense that the symplectic structure $S(J_{A'}\oplus J_{E'})S^t=J_B\oplus J_E$ is maintained for the channel operation. $X_c$ and $Z_c$ are matrices related with the complementary channel in which output is not $B$ but environment $E$, i.e.,  $\gamma \rightarrow X_c\gamma X_c^t + Z_c Z^t_c$.

Finally, in order to compute a coherent information Eq.~(\ref{coh}), we need to know the von Neumann entropy of a Gaussian state. By Williamson theorem~\cite{Williamson}, for any $2n \times 2n$ real symmetric matrix $\gamma$ there exists a symplectic matrix $S$ such that
\begin{equation}\label{Williamson}
S\gamma S^t=\overset{n}{\underset{i=1}{\mathlarger{\mathlarger{\oplus}}}}\lambda_i \mathbb{I}_2.
\end{equation}
The $\lambda_i$ in Eq.~(\ref{Williamson}) can be obtained by computing the eigenvalues of $J\gamma$, as $\pm i \lambda_i$. Then the von Neumann entropy for a Gaussian state having $\gamma$ as its covariance matrix is written as
\begin{align}\label{entropy}
\nonumber H(\rho_\gamma)=&\sum_i \left({\lambda_i+1 \over 2}\right)\tn{log}_2\left({\lambda_i+1 \over 2}\right)\\
&-\left({\lambda_i-1 \over 2}\right)\tn{log}_2\left({\lambda_i-1 \over 2}\right).
\end{align}

\section{main results} \label{main}

\subsection{Gaussian lossy channel}

To show an activation of quantum capacity for Gaussian channels, we begin with similar setting in Ref.~\cite{Smith11}. A special kind of PPT channels, i.e., entanglement binding channels, are needed as activators. On the other side, the `activated channels' are Gaussian lossy channels, i.e., beam splitters with transmissivity $T$. The Gaussian lossy channel is antidegradable for $0 \leq T \leq 0.5$, and degradable for $ 0.5 < T \leq 1$~\cite{Cubitt}. It is known that quantum capacity of antidegradable channel is zero and of degradable lossy channel $\Phi_T$ is written as~\cite{Wolf07}
\begin{equation}\label{lossy}
\mathcal{Q}(\Phi_T)=\tn{max}\left\{ 0,\tn{log}_2 T - \tn{log}_2 (1-T) \right\},
\end{equation}
when $ 0.5 <T \leq 1$.

We now consider the combined channel that consists of the PPT channel and the lossy channel with a particular input state as seen in Fig.~\ref{fig1}, where we use the same PPT channel and input state as those in Ref.~\cite{Smith11}. Explicitly, our three-mode covariance matrix of the input state is
\begin{widetext}
\begin{equation}
\gamma_{in}=\left(
\begin{array}{cccccc}
 \frac{x^4+1}{2 x^2} & 0 & 0 & 0 & \frac{\left(x^4-1\right) \left(y^2-1\right)}{4 x^2 y} & 0 \\
 0 & \frac{x^4+1}{2 x^2} & 0 & 0 & 0 & \frac{\left(x^4-1\right) \left(y^2-1\right)}{4 x^2 y} \\
 0 & 0 & \frac{x^4+1}{2 x^2} & 0 & \frac{\left(x^4-1\right) \left(y^2+1\right)}{4 x^2 y} & 0 \\
 0 & 0 & 0 & \frac{x^4+1}{2 x^2} & 0 & -\frac{\left(x^4-1\right) \left(y^2+1\right)}{4 x^2 y} \\
 \frac{\left(x^4-1\right) \left(y^2-1\right)}{4 x^2 y} & 0 & \frac{\left(x^4-1\right) \left(y^2+1\right)}{4 x^2 y} & 0 & \frac{\left(x^4+1\right) \left(y^4+1\right)}{4 x^2 y^2} & 0 \\
 0 & \frac{\left(x^4-1\right) \left(y^2-1\right)}{4 x^2 y} & 0 & -\frac{\left(x^4-1\right) \left(y^2+1\right)}{4 x^2 y} & 0 & \frac{\left(x^4+1\right) \left(y^4+1\right)}{4 x^2 y^2} \\
\end{array}
\right),
\end{equation}
\end{widetext}
where $x$ and $y$ are appropriate squeezing parameters of the input state, and the following relation should hold for a thermal state input mode $A$, 
\begin{equation}
\bar{n}=\frac{x^4 y^4+x^4-4 x^2 y^2+y^4+1}{8 x^2 y^2},
\end{equation}
where $\bar{n}$ is the average photon number of the thermal state on mode $A$.

The PPT channel used in Ref.~\cite{Smith11} can be represented as
\begin{equation}
X=\begin{pmatrix}\sqrt{2} & 0 & 1 &0 \\ 0 & -\sqrt{2} & 0 & 1 \\ -1 & 0 & 0 & 0 \\ 0 & -1 & 0 & 0 \end{pmatrix}, ~~
Y=\begin{pmatrix}2 & 0 & -\sqrt{2} &0 \\ 0 & 2 & 0 & \sqrt{2} \\ -\sqrt{2} & 0 & 2 & 0 \\ 0 & \sqrt{2} & 0 & 2 \end{pmatrix}.
\end{equation}
\\
This channel satisfies the PPT condition, i.e., $Y+i(J+XJX^t) \geq 0$, and it is indeed an entanglement binding channel since the PPT non-distillability criterion is satisfied, $\gamma_{AB}+i(J_A \oplus -J_B) \geq 0$~\cite{Werner}.

\begin{figure}[!t]
\includegraphics[width=8.5cm]{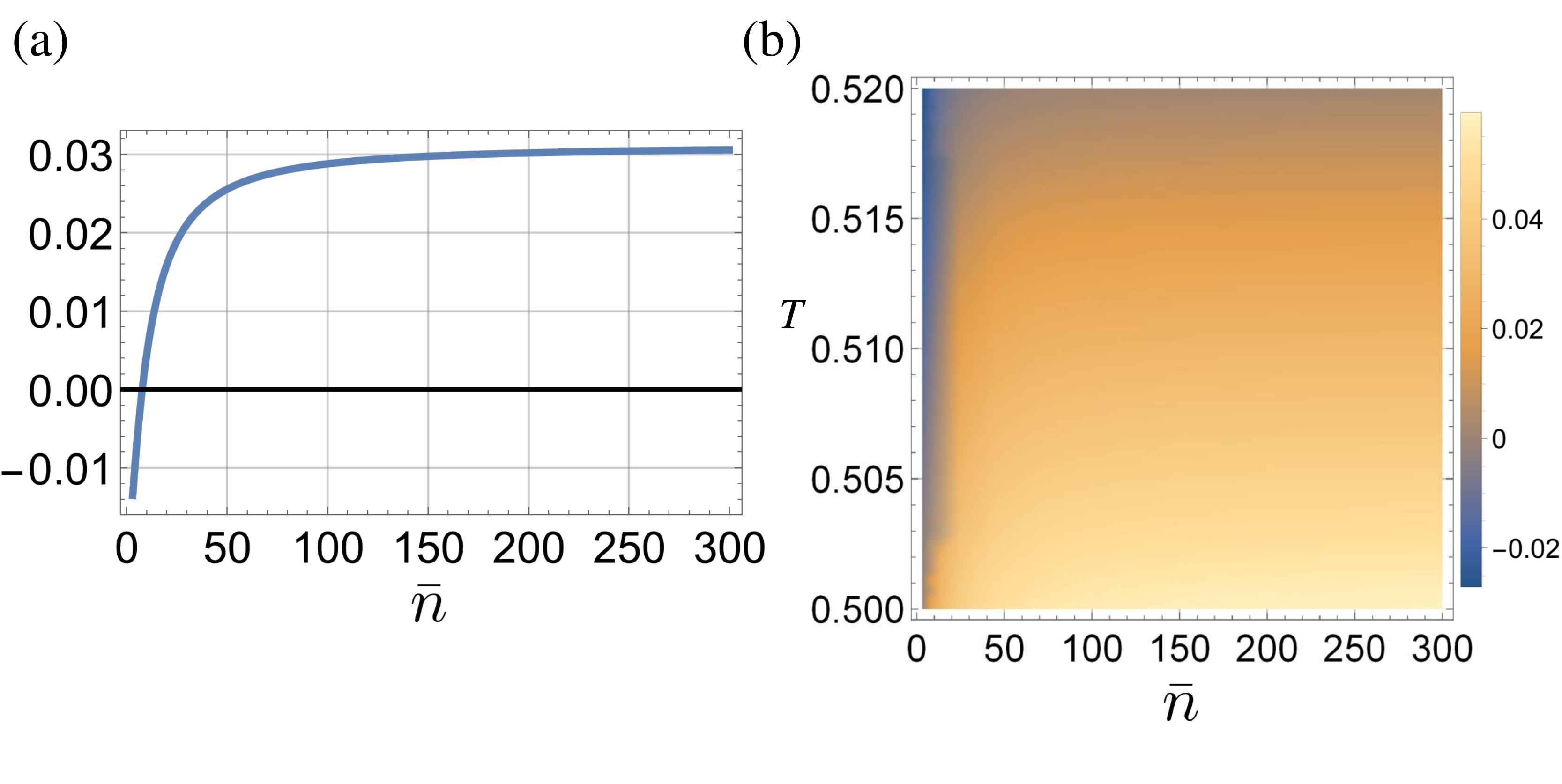}
\caption{Difference $I_c-{\cal{Q}}(\Phi_T)$ as function of input state photon number $\bar{n}$ and transmissivity $T$ of the channel. (a) Case ($T=0.51$); The asymptotic value is $\sim$0.03 bits. (b) Contour diagram of activation with varying $T$; A higher activation appears to occur when $T$ is close to 0.5 and input power is strong.} 
\label{fig2}
\end{figure}

Now we can calculate the coherent information of the combined channel via Eq.~(\ref{entropy}) and the difference between the coherent information and the known quantum capacity in Eq.~({\ref{lossy}), which indeed shows the activation. We plot these results in Fig.~\ref{fig2}. The difference $I_c - \mathcal{Q}(\Phi_T)$ becomes bigger as power of input state increases, or as the transmissivity approaches the superactivation case ($T=0.5$). In addition, the activation occurs when $T\lesssim 0.52$, this means that our PPT channel appears to activate the channels close to the 50:50 beam splitter.

\begin{figure}[!t]
\includegraphics[width=5.5cm]{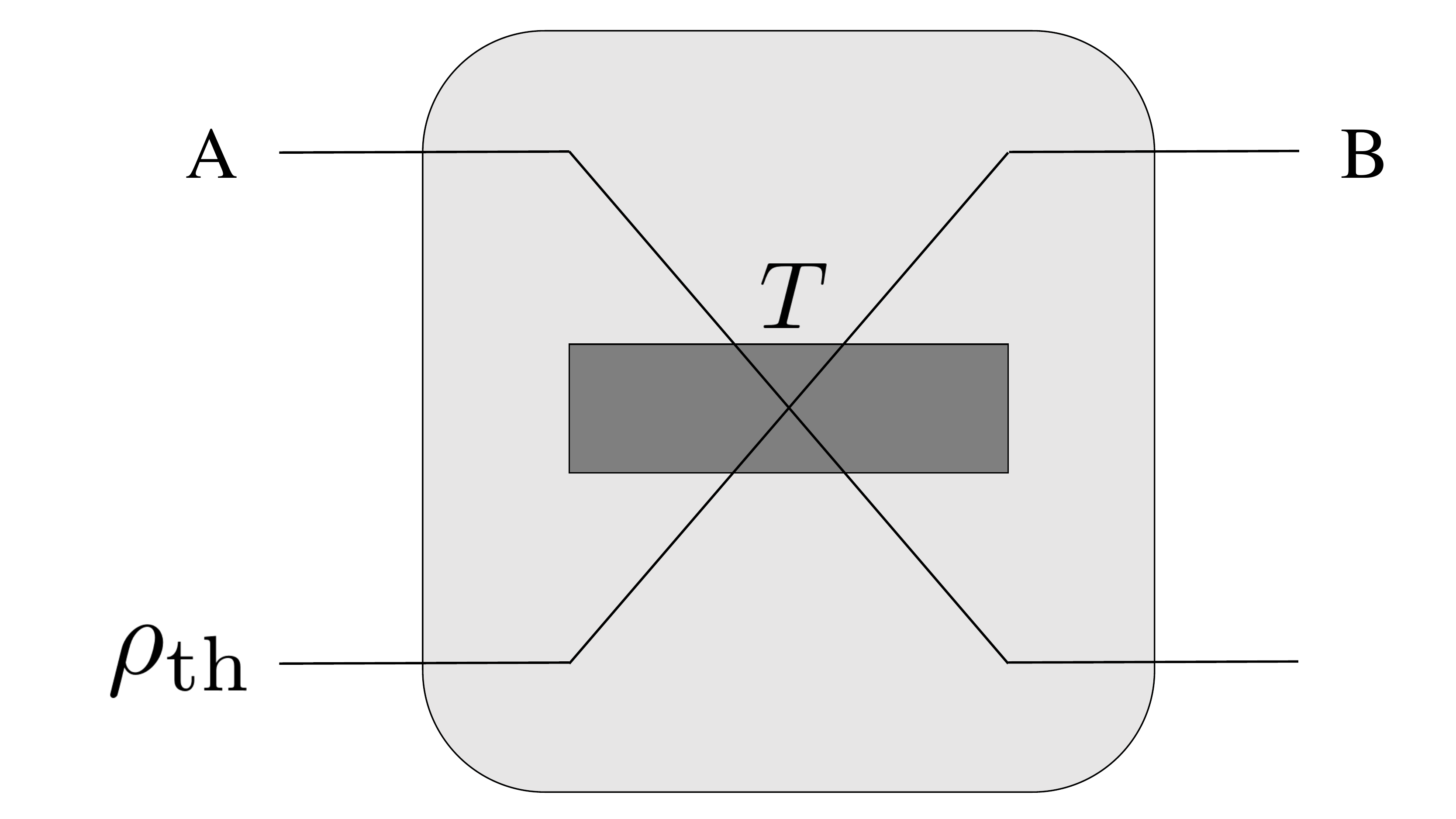}
\caption{A schematic diagram of a thermal attenuator channel. Input mode A is mixing with the thermal state $\rho_{\tn{th}}$. Any Gaussian lossy channel is a special case of this channel having zero temperature.}
\label{fig3}
\end{figure}

\subsection{Gaussian thermal attenuator channel}

A Gaussian thermal attenuator is the general version of Gaussian lossy channels (Fig.~\ref{fig3}), mixing with a thermal Gaussian state instead of the vacuum state. If the channel acts on the input covariance matrix $\gamma$, the output covariance matrix is obtained as
\begin{equation}
\Phi_{T,\bar{N}}(\gamma)=T\gamma+(1-T)(2\bar{N}+1)\mathbb{I}_2,
\end{equation}
where $T$ is the transmissivity of the channel and $\bar{N}$ is the average photon number of the mixing state $\rho_{\tn{th}}$~\cite{ther}. Then the $X_{\tn{th}}$ and $Y_{\tn{th}}$ matrices of this channel are $X_{\tn{th}}=\sqrt{T}\mathbb{I}_2,~Y_{\tn{th}}=(1-T)(2\bar{N}+1)\mathbb{I}_2$, and they satisfy the condition of Gaussian channels, $Y_{\tn{th}}+i(J-X_{\tn{th}}JX_{\tn{th}}^t) \geq 0$.

Unfortunately, the quantum capacity value of this channel has not yet been known, but according to a very recent result in Ref.~\cite{upper}, there is a narrow upper bound on the quantum capacity of this channel, which is tighter than previously known upper bounds~\cite{upper1,upper2,upper3,upper4, upper5} when temperature of the thermal state is low, and the transmissivity is close to 0.5. This is exactly our case, so we can examine whether the PPT channel we consider can activate this channel. The newly found upper bound on the quantum capacity for the thermal attenuator channel $\Phi_{T,\bar{N}}$ can be seen from the inequality~\cite{upper},
\begin{equation}
{\cal{Q}}(\Phi_{T,\bar{N}})\leq \tn{max}\left\{0,\tn{log}_2\frac{\bar{N}(1-T)-T}{(1+\bar{N})(T-1)}\right\},
\end{equation}
when $T \geq 0.5$ and $T>(1-T)\bar{N}$ for a non-entanglement-breaking attenuator condition~\cite{ther}. By using the same input state we can obtain that the coherent information of the combined channel exceeds the upper bound of the channel in a specific parameter region.

In Fig.~\ref{fig4} (a), we plot the difference $I_c-{\cal{Q}}(\Phi_{T,\bar{N}})$ as a function of transmissivity $T$ of the channel and input state photon number $\bar{n}$. It can be seen that the activation occurs when the transmissivity $T$ is slightly above $0.5$ and the average photon number of the thermal attenuator $\bar{N}$ is small. We compare this result with the previous case of the lossy channel in which the exact quantum capacity is known, in Fig.~\ref{fig4} (b). It turns out that the lower bound of the amount of activation of the  thermal attenuator is bigger than that of the lossy channel when $T=0.51$ and $\bar{N}=0.01$. This result looks obviously counterintuitive, since a thermal attenuator with non-zero temperature makes bigger noise through the channel than the zero temperature attenuator. 


\begin{figure}[!t]
\includegraphics[width=8.5cm]{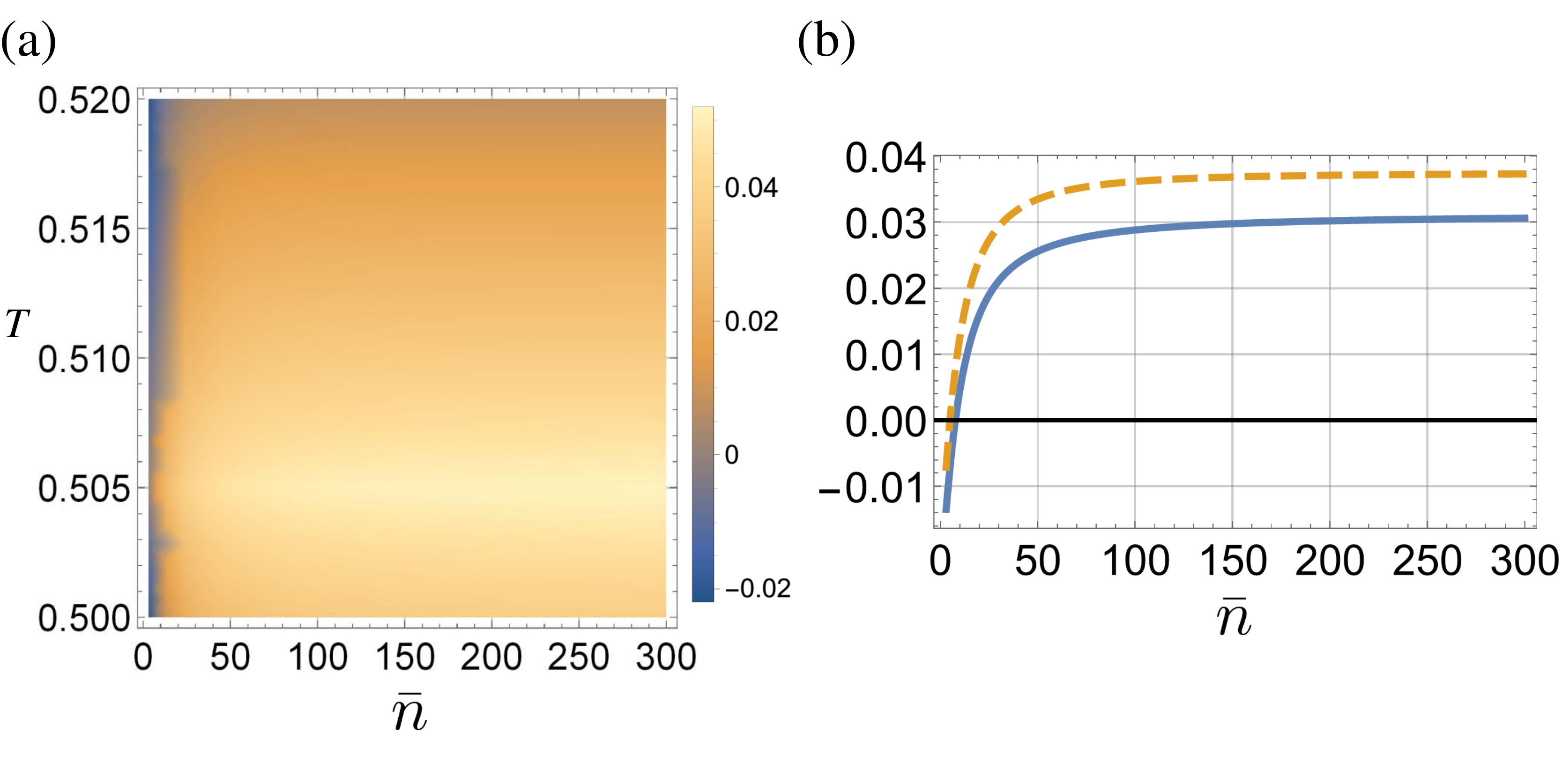}
\caption{(a) Difference between the coherent information of the combined channel and the upper bound on the quantum capacity of the thermal attenuator, $I_c-{\cal{Q}}(\Phi_{T,\bar{N}})$ when $\bar{N}=0.01$. (b) Comparing the thermal attenuator (dashed) with the lossy channel (solid) when $T=0.51$ and $\bar{N}=0.01$.  }
\label{fig4} 
\end{figure}

\section{conclusion}\label{conclusion}

Superactivation is a special case of the more general feature, activation. In this work, we show that the PPT channel used in superactivation also has a capability of activation for the Gaussian lossy channel. Moreover, this channel can also activate a more general Gaussian lossy channel, the thermal attenuator channel.
In addition, its amount of activation appears to exceed that of the lossy channel for some cases, although it might more degrade the channel input than the lossy channel does. Hence, this interesting phenomenon could be helpful to overcome noise of Gaussian channels in more implementable settings.

One can raise a question on whether this activation can be improved. This is crucial because more activation means more robustness to errors of channels, which is related to one of the most important fundamental goals of communication with channels. In this sense, we can consider an energy-constrained quantum capacity of the channel because this is more plausible for physical realizations. In Ref.~\cite{upper4}, several upper bounds of the energy-constrained quantum capacity of a thermal attenuator have been calculated, so we can check the activation effect for those bounds~\footnote{We have seen the fact that such an activation also occurs in this case, but the difference between the coherent information of the combined channel and the upper bound is smaller than in the unconstrained case, as expected.}. For an application for practical situations, we need to consider more implementable input states and PPT channels by adjusting parameters. 

Finally, another important question is whether superactivation or activation can occur by quantum channels other than the entanglement binding channels. It is still an open problem whether other kind of the PPT channels or the NPT channels have potential for (super)activation. It might reveal the role of the bound entanglement for the channel capacity problems.

\section*{ACKNOWLEDGMENTS}

We would like to thank Mark M. Wilde for valuable comments and Gerardo Adesso for helpful discussion. This work was supported by Basic Science Research Program through the National Research Foundation of Korea (NRF) funded by the Ministry of Education (NRF-2017R1A6A3A01007264) and the Ministry of Science and ICT (NRF-2016R1A2B4014928).

\end{document}